\def\ts     {\thinspace}
\def\kms    {\ifmmode{{\rm \ts km\ts s}^{-1}}\else{\ts km\ts s$^{-1}$}\fi}
\def\msol   {\ifmmode{{\rm M}_{\odot} }\else{M$_{\odot}$}\fi}
\def\lsol   {\ifmmode{L_{\odot}}\else{$L_{\odot}$}\fi}
\def\lfir   {\ifmmode{L_{\rm FIR}}\else{$L_{\rm FIR}$}\fi}

\def\,{\thinspace}
\def \ppm{$\pm$}
\def \fran{J2143-4423}
\def \her{{\it Herschel}}
\def \ppm{$\pm$}
\def \ly{Ly$\alpha$}
\def \ba{{B1}}
\def \bb{{B5}}
\def \bc{{B6}}
\def \bd{{B7}}

\documentclass[twocolumn]{aa} 
\usepackage{graphicx}
\usepackage{epsfig}
\usepackage{lipsum}
\usepackage[switch]{lineno}
\usepackage{txfonts}
\begin{document}


   \title{What Powers Ly$\alpha$ Blobs?\thanks{Herschel (Pilbratt et al.
2010) is an ESA space observatory with science instruments provided by
European-led Principal Investigator consortia and with important
participation from NASA.}}

   \author{Y. Ao \inst{1,2}\thanks{email: yiping.ao@nao.ac.jp}, Y. Matsuda \inst{1}, A. Beelen \inst{3}, C. Henkel \inst{4,5},
R. Cen \inst{6}, C. De Breuck \inst{7},
P. J. Francis \inst{8}, A. Kov$\acute{\rm a}$cs \inst{9}, G. Lagache \inst{10},  M. Lehnert \inst{11}, M. Y. Mao \inst{12,13,14},
K. M. Menten \inst{4}, R. P. Norris \inst{13}, A. Omont \inst{11}, K. Tatemastu \inst{1}, A. Wei\ss\, \inst{4} and Z. Zheng \inst{15}}
   \institute{
   National Astronomical Observatory of Japan, 2-21-1 Osawa, Mitaka, Tokyo 181-8588, Japan
   \and
   Purple Mountain Observatory, Chinese Academy of Sciences, Nanjing 210008, China
   \and
   Institut d$'$Astrophysique Spatiale, B$\hat{a}$t. 121, Universit$\acute{\rm e}$ Paris-Sud, 91405 Orsay Cedex, France
   \and
   MPIfR, Auf dem H\"{u}gel 69, 53121 Bonn, Germany
   \and
   Astron. Dept., King Abdulaziz Univ., P.O. Box 80203, Jeddah 21589, Saudi Arabia
    \and
   Princeton University Observatory, Princeton, NJ 08544, USA
   \and
   European Southern Observatory, Karl Schwarzschild Stra{\ss}e 2, 85748, Garching, Germany
   \and
   Research School of Astronomy and Astrophysics, The Australian National University, Canberra ACT 0200, Australia
   \and
   California Institute of Technology 301-17, 1200 E. California Blvd, Pasadena, CA 91125, USA
   \and
   Aix Marseille Universit$\acute{\rm e}$, CNRS, LAM (Laboratoire d$'$Astrophysique de Marseille) UMR 7326, 13388, Marseille, France
   \and
   Institut d$'$Astrophysique de Paris, CNRS and Universit$\acute{\rm e}$ Pierre et Marie Curie, 98bis Bd Arago, 75014 Paris, France
   \and
   School of Mathematics and Physics, University of Tasmania, Private Bag 37 Hobart, 7001 Australia
   \and
   Joint Institute for VLBI, Postbus 2, 7990 AA Dwingeloo, The Netherlands
   \and
   Australia Telescope National Facility, CSIRO Astronomy and Space Science, PO Box 76, Epping, NSW 1710, Australia
  \and
   Department of Physics and Astronomy, University of Utah, Salt Lake City, UT 84112, USA
}
   \date{}

\authorrunning{Y. Ao et al.}
\titlerunning{What Powers the Ly$\alpha$ Blobs?}

\abstract{
Ly$\alpha$ blobs (LABs) are spatially extended \ly\, nebulae seen at high
redshift. The origin of Ly$\alpha$ emission in the LABs is still unclear and
under debate. To study their heating mechanism(s), we present Australia
Telescope Compact Array (ATCA) observations of the 20~cm radio emission and
\her\, PACS and SPIRE measurements of the far-infrared (FIR) emission towards
the four LABs in the protocluster J2143-4423 at $z$=2.38. Among the four LABs,
\bc\, and \bd\, are detected in the radio with fluxes of 67\ppm17~$\mu$Jy and
77\ppm16~$\mu$Jy, respectively, and \bb\, is marginally detected at 3~$\sigma$
(51\ppm16~$\mu$Jy). For all detected sources, their radio positions are
consistent with the central positions of the LABs. \bc\, and \bd\, are
obviously also detected in the FIR. By fitting the data with different
templates, we obtained redshifts of 2.20$^{+0.30}_{-0.35}$ for \bc\, and
2.20$^{+0.45}_{-0.30}$ for \bd\, which are consistent with the redshift of
the \ly\, emission within uncertainties, indicating that both FIR sources are
likely associated with the LABs. The associated FIR emission in \bc\, and \bd\,
and high star formation rates strongly favor star formation in galaxies as an
important powering source for the \ly\, emission in both LABs. However, the
other two, \ba\, and \bb, are predominantly driven by the active galactic
nuclei or other sources of energy still to be specified, but not mainly by star
formation. In general, the LABs are powered by quite diverse sources of
energy.

\keywords{galaxies: formation -- galaxies:high-redshift --
galaxies:ISM -- galaxies:active -- infrared:galaxies} 

}

\maketitle

\section{Introduction}

High-redshift star-forming galaxies are becoming an important probe of galaxy
formation, reionization and cosmology (Robertson et al. 2010; Shapley 2011).  A
popular method for finding high redshift star forming galaxies is to target
their often bright Ly$\alpha$ emission (Partridge \& Peebles 1967). This
emission can be easily detected in narrow-band imaging surveys, and can be
further confirmed by spectroscopic observations (Hu et al. 1998; Ouchi et al.
2008; Yamada et al.  2012a,b). In addition to discovering numerous Ly$\alpha$
emitters (LAEs), a particular class of objects, also known as "Ly$\alpha$
blobs" (LABs), has been most commonly found in the dense environment of
star-forming galaxies at high redshift, and they are very extended (30 to 200
kpc) and Ly$\alpha$-luminous (10$^{43}$ to 10$^{44}$ erg~s$^{-1}$) (see,
e.g., Francis et al. 1996; Steidel et al. 2000; Palunas et al. 2004; Matsuda et
al. 2004, 2009, 2011; Dey et al. 2005; Saito et al. 2006; Yang et al. 2009,
2010; Erb et al.  2011; Prescott et al. 2012a, 2013; Bridge et al. 2013). In
contrast to the large Ly$\alpha$ nebulae surrounding some high-redshift radio
galaxies (e.g., Reuland et al. 2003; Venemans et al. 2007), these objects do
not always have obvious sources for energy responsible for their strong
emission.

While the LABs' preferential location in overdense environments indicates an
association with massive galaxy formation, the origin of Ly$\alpha$ emission in
the LABs is still unclear and under debate (Faucher-Giguere et al. 2010; Cen \&
Zheng 2013; Yajima et al. 2013). Proposed sources have generally fallen into
two categories: cooling radiation from cold streams of gas accreting onto
galaxies (e.g., Haiman et al. 2000; Dijkstra \& Loeb 2009; Goerdt et al. 2010)
and photoionization/recombination from starbursts or active
galactic nuclei (AGNs) (e.g., Taniguchi \& Shioya 2000; Furlanetto et al. 2005;
Mori \& Umemura 2006; Zheng et al. 2011). Supporting evidence for the cooling
flow scenario comes from those LABs lacking any visible power source (e.g.,
Nilsson et al. 2006; Smith \& Jarvis 2007). Ionizing photons from young stars
in star-forming galaxies and/or AGNs can ionize neutral hydrogen atoms and the
subsequent recombination gives off \ly\, emission. The resonant scattering of
\ly\, photons in the circumgalactic medium makes the emission extended (Geach
et al.  2005, 2009; Colbert et al. 2006, 2011; Beelen et al. 2008; Webb et al.
2009; Zheng et al. 2011; Cen \& Zheng 2013; Overzier et al. 2013). 

Except for cooling flows and photoionization from star-forming galaxies and/or
AGNs, other possible mechanisms, such as galactic super-winds and obscured AGNs,
are also proposed to explain the nature of the LABs (e.g., Ohyama et al. 2003;
Wilman et al. 2005; Colbert et al. 2006; Matsuda et al. 2007). All these
sources of energy may be activated in an environment where violent interactions are
frequent between gas rich galaxies as expected in over-dense regions at high
redshift (Matsuda et al. 2009, 2011; Prescott et al. 2012b; Kubo et al. 2013).

The 110 Mpc filament with 37 LAEs related to the protocluster J2143-4423 at
$z$=2.38 (Francis et al. 1996, 2004; Palunas et al. 2004) is one of the largest
known structures at high redshift, and this field also includes four large
extended LABs with extensions of $\sim$ 50 kpc and above, named \ba, \bb, \bc\,
and \bd. In this paper, we present our deep radio observations and {\it
Herschel} released far-infrared (FIR) data in \fran\, to study the powering
source of these LABs. Throughout this paper, we use a $\Lambda$ cosmology with
$\rm H_{\rm0}$ = 67.3~$\rm \kms\ Mpc^{-1}$, $\rm \Omega_\Lambda$ = 0.685 and
$\rm \Omega_{\rm m}$ = 0.315 (Planck Collaboration XVI 2013), and 1$\arcsec$
corresponds to 8.37~kpc at $z$=2.38.

\section{Observations}
\subsection{ATCA observations}
We observed \fran\, with the Australia Telescope Compact Array
(ATCA)\footnote{The Australia Telescope Compact Array is part of the Australia
Telescope, which is funded by the Commonwealth of Australia for operation as a
National Facility managed by CSIRO.} in its extended configuration 6A. During
the observations from 2009 June 14 to 17, only five out of six antennas were
available. The observations were performed at a central frequency of 1.75 GHz.
We used the Compact Array Broadband Backend (Wilson et al. 2011) in a
wide-band mode, with a total bandwidth of 2~GHz and a channel width of 1~MHz.
The nearby source PKS~2134-470 served as a gain calibrator. Absolute fluxes
were calibrated with the ATCA standard PKS~1934-638. The total observing time
was about 70 hours. 

The data were reduced with the MIRIAD software package. Although the observations
were carried out with a total bandwidth of 2~GHz, the effective bandwidth was
about 489 MHz with a central frequency of 1.51~GHz. We carefully flagged the
channels affected by radio frequency interference (RFI) by checking the
visibility data sorted by time, channels and baselines.
The image was deconvolved with MIRIAD task MFCLEAN, and
task SELFCAL was used to reduce the noise from strong radio continuum sources.
We first created cleaned images in a normal procedure and made model images for
the strong sources. The models were used as inputs for task SELFCAL to perform
self-calibration of visibility data. We ran this cycle for three times, and
then obtained the model images to create the visibility data with
self-calibration, which were used to make the final images. The noise of the
images after applying self-calibration was about one order of magnitude lower
than that without self-calibration. The field of view was about 31 arcmins and
the synthesized beam size was 7.8$\arcsec$$\times$4.8$\arcsec$. The noise was
about 15~$\mu$Jy/beam before applying primary beam correction.

\begin{figure*}[t]
\vspace{-0.0cm}
\centering
\includegraphics[angle=0,width=1.0\textwidth]{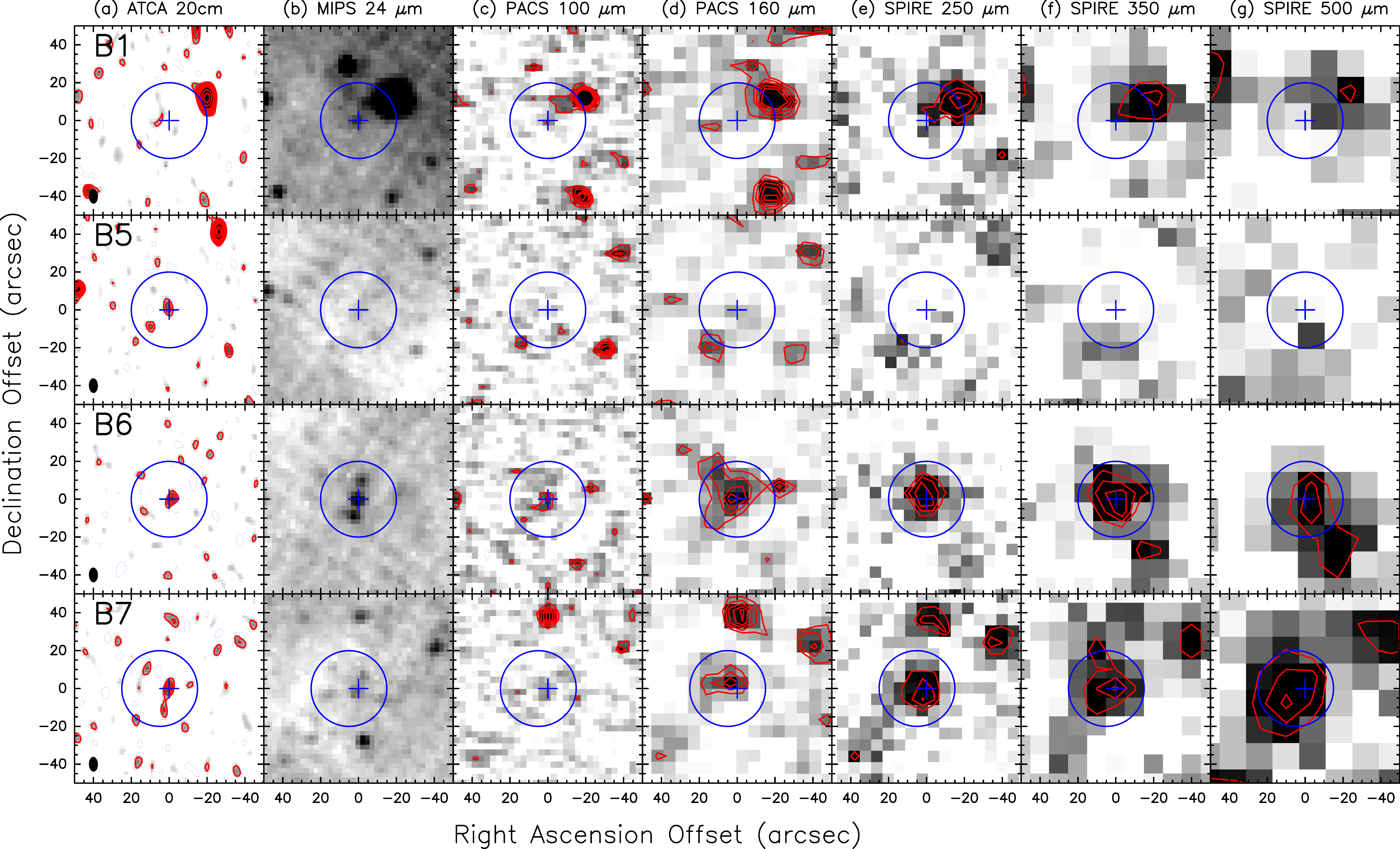}
\vspace{-0.0cm}
\caption{ATCA 20~cm, {\it Spitzer} MIPS 24$\mu$m and \her\, PACS and SPIRE data for the four \ly\, blobs
(LABs) in J2143-4423.  {\bf a) }Contours and gray scale maps of ATCA radio
emission.  The contours are -2, 2, 3, 4, 5 and 6~$\times$ 15~$\mu$Jy
(1~$\sigma$), with a synthesized beam of 7.8$\arcsec$$\times$4.8$\arcsec$,
which is shown in the lower left corner of each panel.  {\bf b) }Gray maps of
{\it Spitzer} MIPS 24~$\mu$m emission (Colbert et al. 2006).  {\bf c-g)
}Contours and gray scale maps of \her\, FIR emission.  The contours are
-2$\sigma$, 2$\sigma$, 3$\sigma$, 4$\sigma$, 5$\sigma$ and 6$\sigma$ (see
$\S$~\ref{obsher} for the noise level of each band). A circle with a diameter
of 40$\arcsec$ is shown in each panel. The circles in
\bd\, are on an off-center position (5$\arcsec$, 0$\arcsec$) to cover most FIR
emission. All sources are centered on the positions of the four LABs (see
Colbert et al.  2006) as shown with plus signs in each panel. All offsets are
relative to the positions of the LABs.}
\label{map} 
\end{figure*}

\subsection{Archival \her\, observations}\label{obsher}
\her\, observations towards \fran\, were carried out with PACS (Poglitsch et
al. 2010) at 100 and 160~$\mu$m and SPIRE (Griffin et al. 2010) at 250, 350 and
500~$\mu$m in 2010 to 2011. \fran\, was imaged in a field size of
15$^\prime$$\times$15$^\prime$ for each band, and the observing time was
$\sim$2.9 hours for PACS (\her\, OD: 686) and $\sim$0.6 hours for SPIRE (\her\,
OD: 558). The level 2.5 product for PACS and the level 2 product for SPIRE from
the pipeline procedures are used for our data analysis. Source photometry
is carried out using DAOphot algorithm in the Herschel Interactive Processing
Environment (HIPE). We apply beam correction, colour correction, aperture
correction for a spectral index of $-$2 and adopt a flux calibration error of 5\%
at PACS bands and 7\% at SPIRE bands as recommended in the PACS and SPIRE
Observer’s Manual. The full width at half power (FWHP) beam sizes are
6.8$\arcsec$ at 100~$\mu$m, 11.4$\arcsec$ at 160~$\mu$m, 17.6$\arcsec$ at
250~$\mu$m, 23.9$\arcsec$ at 350~$\mu$m and 35.2$\arcsec$ at 500~$\mu$m,
respectively.

\section{Results}
\subsection{Radio emission from ATCA observations}
In Fig.~\ref{map}(a) we present the radio continuum emission images at 20~cm
from the ATCA. Among the four LABs, \bc\, and \bd\, are detected with fluxes
of 67\ppm17~$\mu$Jy and 77\ppm16~$\mu$Jy, respectively, and \bb\, is marginally
detected at 3~$\sigma$ (51\ppm16~$\mu$Jy). For all detected sources, their
positions are consistent with the central positions of the LABs.  Only \ba\, is
not detected by the observations. 

\subsection{FIR emission from \her\, observations}
All four LABs are observed with \her\, PACS at 100 and 160~$\mu$m and SPIRE at
250, 350 and 500~$\mu$m, and the images are shown in Fig.~\ref{map}(c-g). The
observed flux densities are calculated for the areas within the blue circles as
shown in Fig.~\ref{map} and are listed in Table~\ref{tab1}. \ba\, is not
detected but contaminated by a nearby strong source about 20$\arcsec$ in the
north-west,  which is the background QSO LBQS2138-4427 at $z$\,=\,3.2 (Francis \&
Hewett 1993), and its emission features at different FIR bands appear to reach
out to \ba\, from this location. There is no FIR counterpart for \bb\, in any
\her\, band.

\begin{center}
\begin{table*}[t]
\centering
\caption{Observational and derived parameters towards the four LABs$^a$}\label{tab1}
\begin{tabular}{ccccccccc}
\hline
Source & 20~cm$^{b}$ & 100~$\mu$m & 160~$\mu$m & 250~$\mu$m & 350~$\mu$m & 500~$\mu$m  
& \lfir$^c$  & M$\rm_{dust}$\\ 
       & [$\mu$Jy]   & [mJy]  & [mJy]  &  [mJy]   &  [mJy] & [mJy]     &  [10$^{12}$\lsol]  & [$10^8$\msol] \\
\hline
B1  & $<$51    &   $<$4.2   &  $<$9.0  &  $<$17.9  &  $<$19.6  &   $<$22.5    &  $<$2.8 & \\
B5  & 51\ppm16 &   $<$2.1   &   $<$11.1  & $<$17.5  & $<$18.7  &   $<$19.8    &  $<$2.5 & \\
B6  & 67\ppm17 &  13.2\ppm3.2   &  53.9\ppm8.0  & 49.7\ppm9.0  & 53.7\ppm10.7  &  36.7\ppm10.3    & 10.0\ppm1.9  & 3.2\ppm0.6 \\
B7  & 77\ppm16 &  12.9\ppm4.0   &  33.5\ppm10.0  & 41.6\ppm7.8  & 48.0\ppm10.6  &  39.2\ppm8.6    & 8.6\ppm2.3  & 5.0\ppm1.0\\
\hline
\end{tabular}
\begin{list}{}{}
\item{$^{\mathrm{a}}$ The wavelengths shown in this table are the redshifted values.}
\item{$^{\mathrm{b}}$ Measured fluxes have been modified by a primary beam correction (less than 15\%).}
\item{$^{\mathrm{c}}$ The total luminosities are calculated between rest frame
wavelengths of 40~$\mu$m to 200~$\mu$m from the dust models (see
$\S$~\ref{dust} for details). The 3~$\sigma$ upper limits are given for undetected sources.} 
\end{list}
\end{table*}
\end{center}

\subsection{Redshifts of the FIR sources}\label{red}
To estimate the redshift of the FIR sources associated with \bc\, and \bd, we
try to fit the data with the SEDs of different templates (Polletta et al.
2007) at different redshifts and find that the starburst templates can well
reproduce the data.  With the observational data and the SEDs of the templates,
the minimum reduced $\chi^2$ value for each redshift can be calculated and the
corresponding probability can be estimated. In this analysis, we include
five \her\, band, APEX 870~$\mu$m data (Beelen et al. 2008), and {\it Spitzer}
MIPS 24~$\mu$m data (Colbert et al.  2006).

\begin{figure*}[t]
\vspace{-0.0cm}
\centering
\includegraphics[angle=0,width=1.0\textwidth]{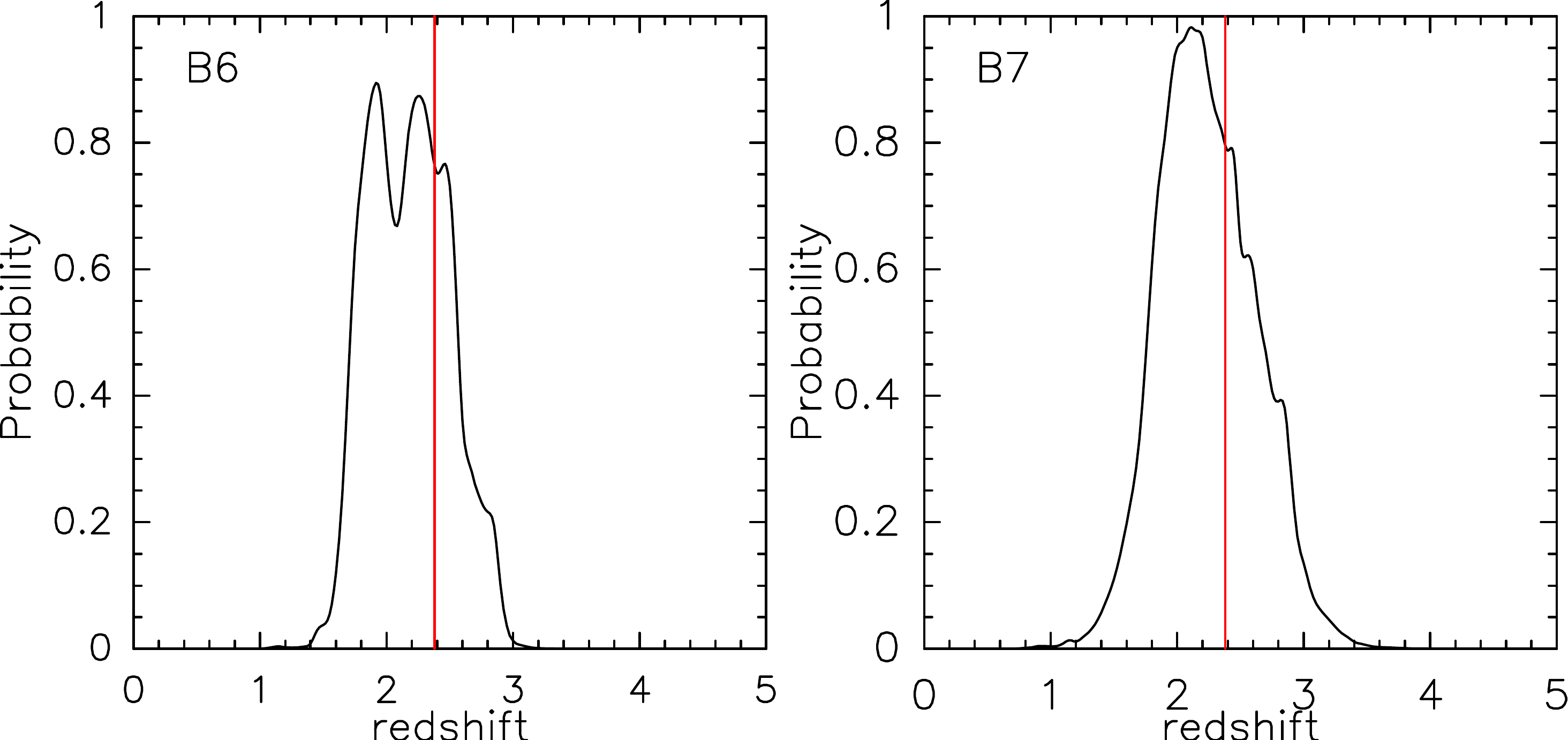}
\vspace{-0.0cm}
\caption{Probability as a function of redshift for \bc\, and \bd. NGC~6240 and
Arp~220 are adopted as the most appropriate starburst templates for B6 and B7, respectively.
A red vertical line denotes a redshift of 2.38 for
\ly\, emission.}
\label{redshift} 
\end{figure*}

Among four typical templates, Arp~220, M~82, Mrk~231 and NGC~6240, we find
that the spectral energy distribution of starburst galaxies NGC 6240 and Arp 220
fit the data best, and Mrk 231 doesn't fit well because it has warm IR emission 
from its AGN which is not really consistent with the data.
Fig.~\ref{redshift} shows the probability distribution
against redshift for both LABs. The estimated redshifts are
2.20$^{+0.30}_{-0.35}$ for \bc\, and 2.20$^{+0.45}_{-0.30}$ for \bd\,
respectively. Considering the uncertainty of this method to determine the
redshifts, both values are consistent with the \ly\, redshift of 2.38 of the
LABs. Adopting the number count study of \her\, sources in Clements et al.
(2010), the probability of finding a 350 $\mu$m source with a flux greater than 40 mJy
within 20 arcsec is 2\%. According for
such a low number density of strong FIR sources and the positional coincidence
of the LABs with strong FIR sources, the FIR sources are very likely associated
with the LABs. Nevertheless, future spectroscopic observations from molecular
lines at millimeter or from forbidden lines at near-infrared will be quite
important to confirm it. In the following sections, the \ly\, redshift of 2.38
will be adopted for the LABs.

\subsection{Dust properties}\label{dust}
\begin{figure*}[t]
\vspace{-0.0cm}
\centering
\includegraphics[angle=0,width=1.0\textwidth]{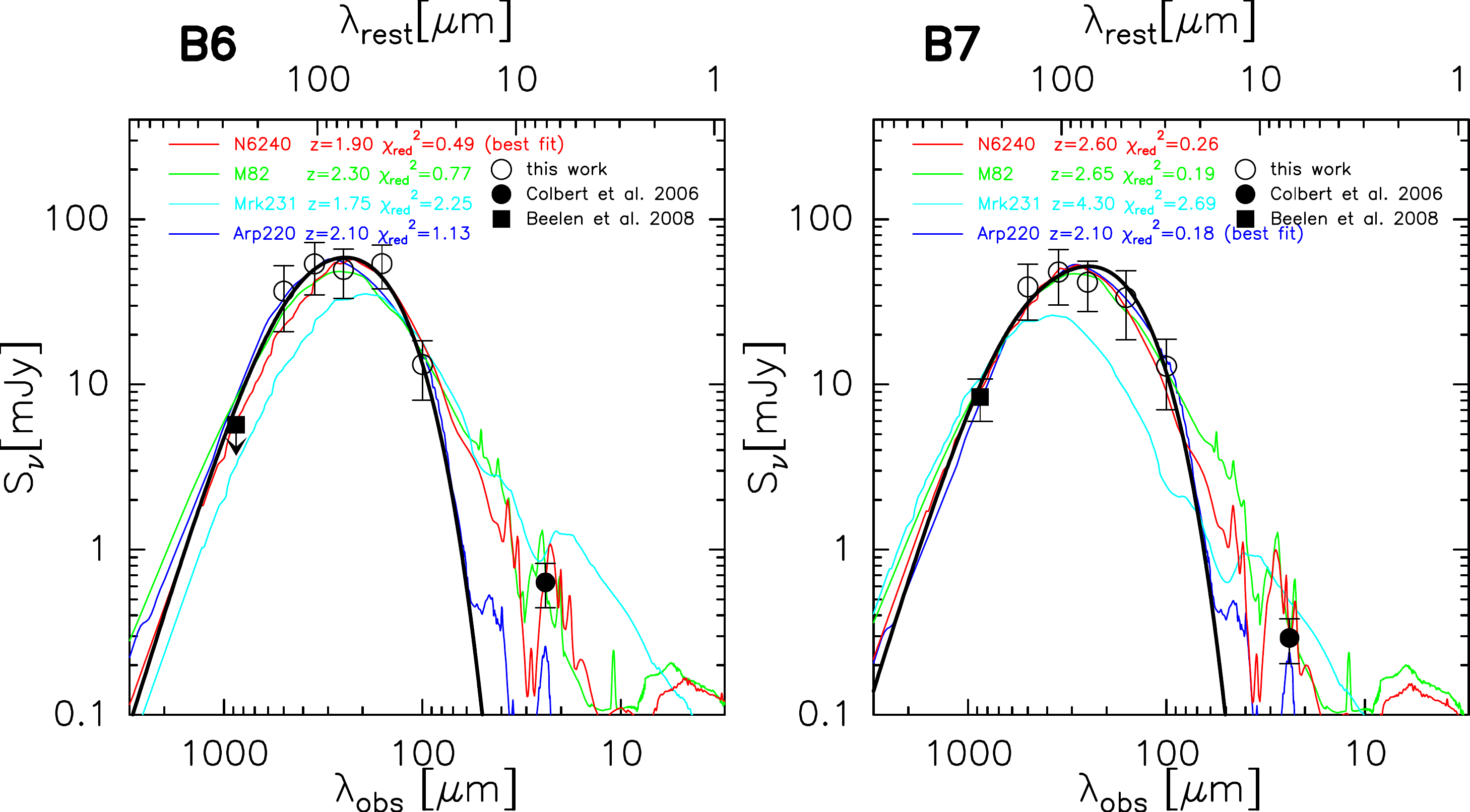}
\vspace{-0.0cm}
\caption{Single-component dust models for B6 and B7 (a redshift of 2.38 is
adopted). The black solid lines show the thermal dust continuum emission
of the 70~K dust components for both \bc and \bd. The open circles represent
the measurements at five \her\, bands in this paper and the filled circles
indicate the flux densities at 24~$\mu$m (Colbert et al.  2006).  The filled
square denotes the flux density (or its upper limit) at 870~$\mu$m, taken from
Beelen et al. (2008). The wavelengths at the rest frame are labelled on the
top. For the single-component dust models adopted in the figure (see
$\S~\ref{dust}$ for details of the dust models.), the $\chi^2$ values are 1.1
for \bc\, and 1.0 for \bd, respectively. In $\S~\ref{red}$, four typical
starburst templates, NGC~6240, M~82, Mrk~231 and Arp~220 (Polletta et al.
2007), are adopted to estimate the redshifts for \bc\, and \bd, and their best
fits are overlaid in colored lines.
}
\label{sed} 
\end{figure*}

For \bc\,and \bd, we have included the measurements from the five \her\, bands
as well as the 870~$\mu$m data taken from Beelen et al. (2008)
in the dust continuum analysis using a single-component dust model as described in
Wei\ss\ et al. (2007). 
{\it Spitzer} MIPS 24~$\mu$m data (Colbert et al. 2006) are not used in the
model fitting because they are strongly affected by PAH features, but are shown
in Fig.~\ref{sed} to allow for a better comparison with overlaid templates.
We find a dust temperature, $T\rm_{dust}$, of
70\ppm5 K and a dust mass, M$\rm_{dust}$, of (3.2\ppm0.8)$\times$10$^8$ M$\rm
_\odot$ for \bc, and $T\rm_{dust}$\,=\,70\ppm5 K and
M$\rm_{dust}$\,=\,(5.0\ppm1.0)$\times$10$^8$ M$\rm _\odot$ for \bd, respectively. 
The implied FIR luminosities are \lfir\, = (10.0\ppm1.9)$\times$10$^{12}$\,\lsol\,
for \bc, and \lfir\, = (8.6\ppm2.3)$\times$10$^{12}$\,\lsol\, for \bd,
respectively, where \lfir\, is integrated from 40~$\mu$m to 200~$\mu$m in the
rest frame. The upper \lfir\,limits for both \ba\, and \bb\, are
$\sim$2.5$-$2.8$\times$10$^{12}$\,\lsol.

\subsection{Star formation rates}
Here we derive the star formation rates from the \ly, far-infrared and
radio luminosities.  To estimate the star formation rate (SFR) from the \ly\,
luminosity, we first assume that star formation (SF) powers the observed
\ly\, flux.  We use an unreddened Ly$\alpha$/H$\alpha$ ratio of 8:1 and the
conversion factor between H$\alpha$ luminosity and SFR (Kennicutt 1998),
yielding SFR($\rm{Ly\alpha}$)/(\msol/yr)\,=\,$L_{\rm Ly\alpha}$/(10$^{42}$ erg
s$^{-1}$). This provides a lower limit because the extinction of \ly\, emission
caused by dust will largely reduce the observed \ly\, luminosity. With the FIR
luminosity derived from \her\,data, we can estimate the SFR by using the
relation SFR($L_{\rm FIR}$)/(\msol/yr)\,=\,$1.7\times$$L_{\rm FIR}$/(10$^{10}$
\lsol) (Kennicutt 1998). If the observed radio emission, with a rest wavelength
of 6~cm, is dominated by free-free emission in H{\small II} regions, one can
also relate the SFR by the relation SFR($L_{\rm
1.4~GHz}$)/(\msol/yr)\,=\,5.52$\times$10$^{-22}$~$L_{\rm 1.4~GHz}$/(W Hz$^{-1}$)
(Bell 2003). The radio luminosity at 1.4 GHz at the rest frame can be estimated
from the observed flux at 1.51 GHz by assuming a relation
$S \propto \nu^\alpha$, where S is the flux density and the typical spectral
index $\alpha$ of $-$0.8 is commonly adopted for the SMGs (e.g., Ivison et al.
2010). These values are listed in Table~\ref{tab2}.

\begin{center}
\begin{table}[h]
\centering
\caption{Derived star formation rates towards the four LABs}\label{tab2}
\begin{tabular}{ccccc}
\hline
Source &  SFR(${L_{\rm FIR}}$) & SFR($L_{\rm 1.4GHz}$) & log $L_{\rm Ly\alpha}$$^a$  &  SFR($\rm{Ly\alpha}$) \\ 
       &  [\msol/yr] & [\msol/yr] &  [ergs~s$^{-1}$] & [\msol/yr]  \\
\hline
B1  &  $<$480 & $<$1090 & 43.9 & 79 \\
B5  &  $<$430 & 1090\ppm340 & 43.8 & 63 \\
B6  &  1700\ppm320    & 1430\ppm360 & 43.8 & 63 \\
B7  &  1460\ppm390    & 1650\ppm340 & 43.5 & 32  \\
\hline
\end{tabular}
\begin{list}{}{}
\item{$^{\mathrm{a}}$ The \ly\,luminosities are adopted from Colbert et al. (2006).}
\end{list}
\end{table}
\end{center}

\section{Discussion and Conclusions}

A high detection rate of radio emission (three out of four) around LABs
suggests that most LABs do not originate from cooling radiation. Instead,
photoionization from starbursts and/or AGNs may power the LABs in most cases.
The high rate of FIR detections (two out of four) points to a star-formation
origin of the LABs. 
The SEDs of \bc\, and \bd\, can be well described by starburst dominated
templates, as shown in Fig.~\ref{sed}, further supporting \ly\, emission
related to the SF in the LABs. In \bc\, and \bd, the SFRs derived from \ly\,
fluxes are far below those estimated from FIR luminosities (Table~\ref{tab2}).
This suggests that the dust indeed greatly reduces the measured \ly\, flux.
Comparing the different SFRs, the dust absorption optical depth of the \ly\,
emission becomes $\sim$3.1$-$3.6. The SFRs estimated from the FIR and radio
luminosities are comparable, indicating that the radio emission is dominated by
SF, not by AGNs. The energetic starbursts can provide enough ionizing photons
to ionize neutral hydrogen atoms in the interstellar medium (ISM), and each
subsequent recombination has a probability of $\sim$ 2/3 of ending up as a
Ly$\alpha$ photon (Partridge \& Peebles 1967).  After escaping the galaxy's
ISM, these Ly$\alpha$ photons can be resonantly scattered by neutral hydrogen
atoms in the intergalactic medium (IGM), which tends to make the Ly$\alpha$
emission extended (Zheng et al.  2011). 

Cen \& Zheng (2013) propose an SF-based model and predict that LABs at high
redshift correspond to protoclusters containing the most massive galaxies and
cluster halos in the early universe as well as ubiquitous strong infrared
sources undergoing extreme starbursts. 
This may be supported by the multiple Spitzer/MIPS sources detected in both
LABs (see Fig~\ref{map}(b), Colbert et al.  2006, 2011).  Indeed, Prescott et
al. (2012b) suggest that LABs may be the seeds of galaxy clusters by resolving
the galaxies within a LAB at $z$\,=\,2.7. The strong FIR emission and the
inferred high SFRs support the presence of a strong starburst in both \bc\, and
\bd. However, AGN-dominated templates like Mrk~231 can not well reproduce
the data (see $\S$~\ref{red}), suggesting that the SF instead of AGN may power
the \ly\, emission in both LABs. The model also predicts that the most
luminous FIR source in each LAB is likely representing the gravitational center
of the protocluster. Fig.~\ref{map}(c-g) shows that the FIR emission indeed
peaks in the centers of \bc\, and \bd. The radio continuum emission is detected
exclusively in the centers, which suggests that the source with most luminous
FIR emission (therefore highest SFR) is in the gravitational center of each
LAB. Another very important prediction of this model is that the \ly\, emission
from photons that escape the galaxy are expected to be significantly polarized,
which has been for the first time confirmed towards LAB1 in the SSA22 field by
Hayes et al. (2011), supporting models with central power sources. Adopting a
gas-to-dust mass ratio of 150 and the SFRs estimated above, the timescales of
\bc\, and \bd\, are relatively short ($\sim$100 Myr), which is much shorter
than the galaxy building timescale. Note that this timescale is a lower limit
because (1) the LABs may have been alive for a while now, and (2) additional
gas may be continuously accreted. In any case, the LABs are visible only during
a short time interval during the lifetime of their parent clusters.

Note that the so-called ``SF-based model'' proposed by Cen \& Zheng (2013) also
includes AGN powering or any central powering. The morphologies of the
\ly\,emission of the four LABs are quite different (Palunas et al. 2004):
\ba\,and \bb\, have core-like structures, while \bc\,and \bd\, are
characterized by diffuse and extended emission with physical sizes of
$\sim$60-70~kpc. The latter may be driven by multiple sources as suggested
by the MIPS data and are consistent with the SF-based model.  There is no
clear FIR emission detected around \ba\,and \bb. Therefore, the \ly\, emission
in both LABs is unlikely predominantly triggered by SF. Overzier et al.  (2013)
conclude that in \ba\, the photoionization from an AGN is the main driver of
\ly\, emission. However, Francis et al. (2013) shows that the observed \ly\,
emission in \ba\, is of complex origin, dominated by the sum of the emission
from the sub-haloes where the cold gas is being lit up most likely by a
combination of tidally triggered star formation, bow shocks, resonant
scattering of \ly\, from the filament collisions and tidal stripping of the
gas. In \bb\, radio emission is tentatively detected and therefore the AGN may
also power the \ly\, emission. Among the four LABs in \fran, two of them, \bc\,
and \bd, are mainly driven by SF. However, the other two LABs, \ba\, and \bb,
without clear FIR detection, are predominantly driven by the AGNs or other
sources of energy still to be specified, but not mainly by star formation. We
thus conclude that LABs must be powered by quite diverse sources of energy.

With its high angular resolution and superb sensitivity, future observations
with the Large Atacama Millimeter Array (ALMA) will reveal more details about
the nature of LABs such as testing the predictions of models where the
ionization is provided by intense star formation and confirming the
significantly polarized dust emission at mm/submm wavelength.

\begin{acknowledgements}
We thank the anonymous referee for valuable comments that improved this manuscript. 
Y.A. acknowledges partial support by NSFC grant 11373007 and Youth Innovation Promotion Association CAS.
R.C. is supported in part by NASA grant NNX11AI23G.
Y.M. acknowledges support from JSPS KAKENHI Grant Number 20647268. 
ZZ was partially supported by NSF grant AST-1208891 and NASA grant NNX14AC89G.
This research has made use of NASA's Astrophysical Data System (ADS).

PACS has been developed by a consortium of institutes led by MPE (Germany) and
including UVIE (Austria); KU Leuven, CSL, IMEC (Belgium); CEA, LAM (France);
MPIA (Germany); INAF-IFSI/OAA/OAP/OAT, LENS, SISSA (Italy); IAC (Spain). This
development has been supported by the funding agencies BMVIT (Austria),
ESA-PRODEX (Belgium), CEA/CNES (France), DLR (Germany), ASI/INAF (Italy), and
CICYT/MCYT (Spain).

SPIRE has been developed by a consortium of institutes led by Cardiff
University (UK) and including Univ. Lethbridge (Canada); NAOC (China); CEA, LAM
(France); IFSI, Univ. Padua (Italy); IAC (Spain); Stockholm Observatory
(Sweden); Imperial College London, RAL, UCL-MSSL, UKATC, Univ. Sussex (UK); and
Caltech, JPL, NHSC, Univ. Colorado (USA). This development has been supported
by national funding agencies: CSA (Canada); NAOC (China); CEA, CNES, CNRS
(France); ASI (Italy); MCINN (Spain); SNSB (Sweden); STFC, UKSA (UK); and NASA
(USA).
\end{acknowledgements}

\end{document}